\documentclass[runningheads]{llncs}

\usepackage{graphicx}
\usepackage{xcolor}
\usepackage{hyperref}
\usepackage[T1]{fontenc}
\usepackage{booktabs}
\usepackage{array}

\begin{document}

\title{LearnLens: An AI-Enhanced Dashboard to Support Teachers in Open-Ended Classrooms}
\titlerunning{LearnLens: An AI-Enhanced Dashboard to Support Teachers}



\author{Namrata Srivastava\thanks{These authors contributed equally to this work.} \and Shruti Jain\samethanks \and Clayton Cohn \and Naveeduddin Mohammed \and Umesh Timalsina \and Gautam Biswas}
\makeatletter
\newcommand{\samethanks}[1][\value{footnote}]{\footnotemark[#1]}
\makeatother

\institute{Vanderbilt University, USA \\
\email{\{namrata.srivastava, shruti.jain, clayton.a.cohn, naveeduddin.mohammed, umesh.timalsina, gautam.biswas\}@vanderbilt.edu}
}

\authorrunning{N. Srivastava et al.}
\maketitle              
\begin{abstract}
Exploratory learning environments (ELEs), such as simulation-based platforms and open-ended science curricula, promote hands-on exploration and problem-solving but make it difficult for teachers to gain timely insights into students' conceptual understanding. This paper presents LearnLens, a generative AI (GenAI)-enhanced teacher-facing dashboard designed to support problem-based instruction in middle school science. LearnLens processes students' open-ended responses from digital assessments to provide various insights, including sample responses, word clouds, bar charts, and AI-generated summaries. These features elucidate students' thinking, enabling teachers to adjust their instruction based on emerging patterns of understanding. The dashboard was informed by teacher input during professional development sessions and implemented within a middle school Earth science curriculum. We report insights from teacher interviews that highlight the dashboard’s usability and potential to guide teachers' instruction in the classroom.
\end{abstract}


\section{Introduction and Background}

Exploratory Learning Environments (ELEs), such as simulation-based platforms and game-based settings, provide students with opportunities to explore diverse problem spaces and experiment with various learning strategies. These environments prioritize conceptual understanding and the development of critical thinking skills over the mere acquisition of procedural knowledge \cite{mavrikis_intelligent_2019}. While ELEs offer valuable opportunities for students to gain a deeper understanding of a domain and develop problem-solving skills by encouraging them to make decisions, test ideas, and construct meaning through hands-on exploration, they also pose significant challenges for students and teachers. 

In real classrooms, it can be challenging for teachers to gain a comprehensive understanding of each student’s progress, thought processes, and misconceptions, especially when students are working independently or in small groups on open-ended tasks~\cite{dillenbourg_design_2013,mavrikis_design_2016}. Since teachers can only interact with a limited number of students at a time, much of the students’ thinking remains hidden from view. Prior research suggests that utilizing interactive technologies to make students' ideas more visible can help instructors provide more effective guidance, thereby enhancing student learning \cite{moed-abu_raya_teachers_2024,zhai_design_2024,hutchins_using_2023}. However, there is still a lack of scalable and pedagogically effective tools that assist teachers in interpreting open-ended responses in online learning environments in a timely manner.

Recent advances in generative AI (GenAI), particularly large language models (LLMs), offer significant opportunities for analyzing open-ended student responses. This includes tasks such as grading short answers and essays, reviewing student self-explanations, and identifying misconceptions~\cite{giannakos_promise_2024,yan2024practical}. These models can provide valuable feedback for both formative and summative assessments, making them especially useful in open-ended and exploratory learning environments. GenAI's capability to identify patterns across diverse student responses and offer concise summaries of collective class understanding makes it a valuable tool for instructional decision-making.

A teacher-facing dashboard powered by GenAI provides a promising design model for this kind of support~\cite{zhai_design_2024}. Dashboards that aggregate student-generated data with interactive visualizations and AI-generated insights can more effectively support educators, particularly when co-designed with them~\cite{hutchins_codesigning_2024}.

In this paper, we present LearnLens, a GenAI-supported interactive teacher-facing dashboard that offers multiple layers of insights into students' thinking in an exploratory learning classroom. A key innovation of LearnLens is its AI-generated insights feature, which synthesizes student responses to highlight collective understanding and common misconceptions,  providing teachers with timely, actionable summaries that are often difficult to obtain in real time. The dashboard ingests students' open-ended responses from digital classroom activities (such as check-ins, exit tickets, and formative assessments). It generates interactive visualizations and AI-generated summaries and insights to make students' thinking more visible and interpretable in real time, allowing teachers to adapt their instructions accordingly. 

LearnLens was implemented as a high-fidelity prototype within a middle school Earth science curriculum that integrates science, computing, and engineering concepts together in a 15-lesson curriculum~\cite{basu2022promoting}. The dashboard's features were informed by researcher-teacher interactions during the professional development sessions before running the classroom studies and were designed to address the challenges teachers face when interpreting open-ended responses in real-time. To evaluate the usability and instructional relevance of the dashboard, we conducted semi-structured interviews with two science teachers, focusing on how the dashboard could support their decision-making while teaching.

We present findings from one joint interview of two teachers. The teachers appreciated the dashboard's clear visualizations and modular interface. They also found the GenAI component useful for identifying gaps and misconceptions in students' understanding. Our teacher-facing dashboard aligns with several design principles identified by Poh et al. (2023) \cite{poh_design_2023} for supporting in-class learning. These principles include providing real-time insights to inform instruction, supporting teachers' pedagogical goals, and minimizing the effort needed to interpret and act on the data. 

Overall, this work supports the workshop's goals by exploring how AI, particularly Generative AI (GenAI), can enhance teacher decision-making in exploratory learning environments. It contributes to ongoing efforts to integrate GenAI to support STEM education \cite{price2025generative}.
\section{Designing the LearnLens Dashboard}

\subsection{Study Context: Earth Science Curriculum}
The dashboard was implemented as a high-fidelity prototype within a 15-lesson middle school NGSS-aligned Earth science curriculum that integrates science, computing, and engineering concepts \cite{basu2022promoting}. The curriculum covered topics such as conservation of mass, rainfall, absorption, and runoff, helping students understand the challenges of urban water runoff. Students began the curriculum with science lessons on rainwater absorption by different surface materials, before transitioning to computational modeling, utilizing a block-based Domain-Specific Modeling Language (DSML) \cite{hutchins2020domain}. Finally, students engaged with a computer-based Engineering Design Challenge, where they applied their knowledge to design a schoolyard that minimizes runoff after heavy rainfall.

Daily assessments included exit tickets (N = 8), check-ins (N = 3), and formative assessments (N = 4), administered digitally as Google Forms. Exit tickets and check-ins were ungraded and intended solely to monitor students' learning; exit tickets captured student reflections, vocabulary development, and areas of confusion, while check-ins assessed understanding of key concepts discussed in class. In this paper, we demonstrate the potential of an AI-enhanced dashboard using data from a student check-in conducted after the end of Lesson 1.

\subsection{Dashboard Implementation and Design}
The dashboard components were guided by professional development sessions with expert teachers, who emphasized the need to assess students' understanding of core curriculum concepts at the end of each lesson to adapt their lesson plans. In previous implementations of the curriculum, the researchers were only able to demonstrate limited aspects of students' conceptual understanding (e.g., showing sample responses and word clouds on a PowerPoint slide). 

\begin{figure}[!ht]
    \centering
    \includegraphics[width=1\linewidth]{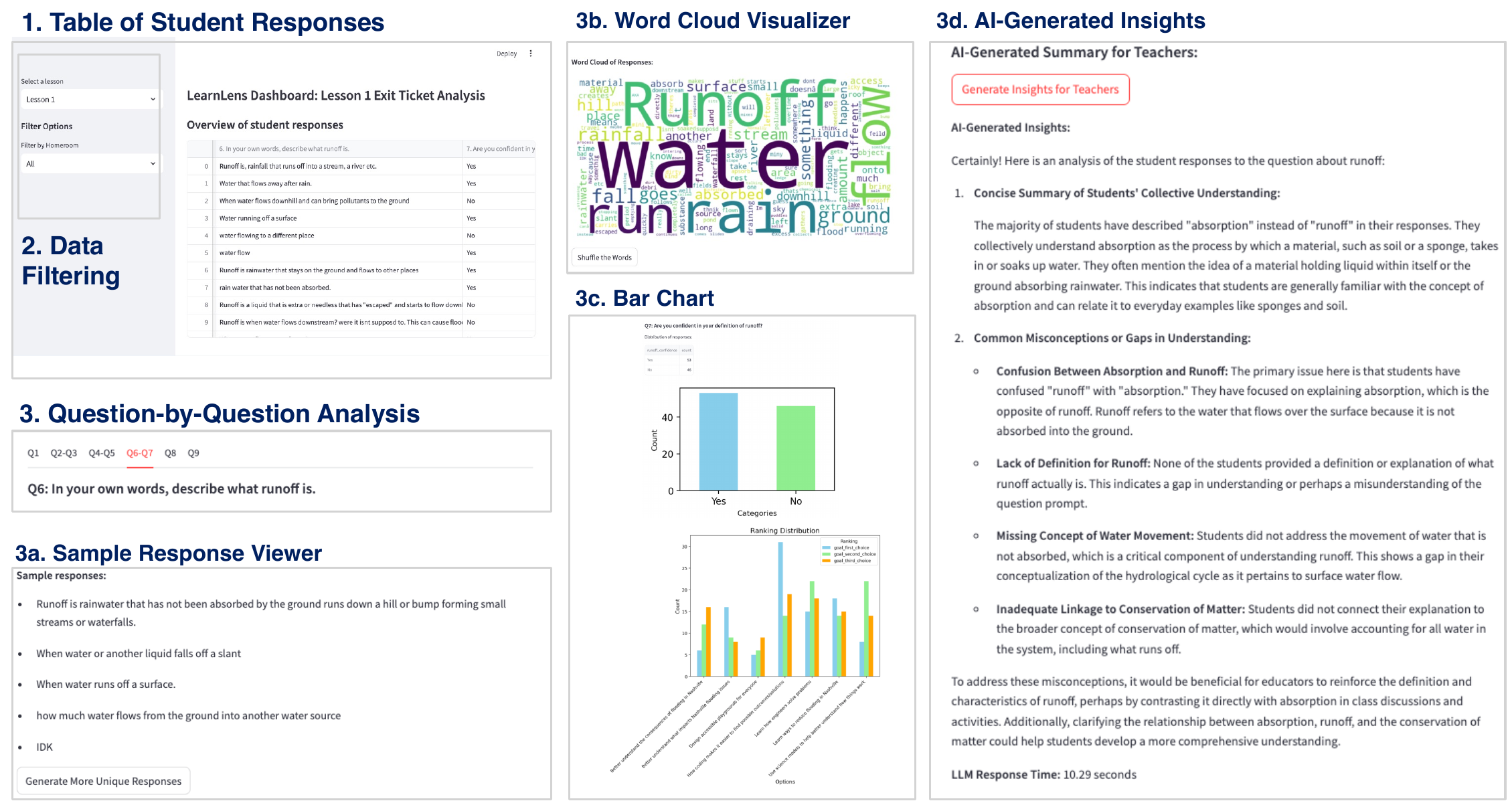} 
    \caption{Features of LearnLens Dashboard}
    \label{fig:dashboard}
\end{figure}

The LearnLens dashboard was designed to build upon these earlier efforts. It was developed using Python to display data online (after the end of each class) and includes the following core features (see Figure \ref{fig:dashboard} for an overview):

\begin{itemize}
    \item \textbf{Table of Student Responses:} This table displays all student responses for an assessment, helping teachers quickly review the data.
    \item \textbf{Data Filtering}: The dashboard features a filter to allow teachers to view the performance of individual class sections by homeroom. 
    
    \item \textbf{Question-by-Question Analysis:} Each question appears in its own tab to reduce information overload. For every question, the dashboard includes:
    \begin{itemize}
        \item \textit{Sample Response Viewer}: This function allows teachers to scroll through a range of student responses. 
        Only five responses are shown at a time, with the option to view more by clicking the ``Generate More Unique Responses'' button.
        \item \textit{Word Cloud Visualizer}: A word cloud of students' responses was generated using the Python WordCloud package. The visualization highlights common vocabulary and themes across student responses. It also has the option to shuffle the responses.
        \item \textit{Bar Chart (for non-open-ended questions)}: A bar chart provided a quick summary of how students responded to multiple-choice or checkbox items.
        \item \textit{AI-Generated Insights}: LLM model summarized students'  understanding and identified common misconceptions in their understanding. 
    \end{itemize}
\end{itemize}

Among these, the AI-generated insights feature was the most novel and required careful refinement. We describe its development in the next section.

\subsection{Development of the GenAI Component}

The AI-generated insights in LearnLens were powered by the OpenAI GPT-4o model (version gpt-4o-2024-08-06)\footnote{\url{https://openai.com/index/hello-gpt-4o/}}. The GenAI component was designed to (1) provide a concise summary of students' collective understanding and (2) identify common misconceptions or gaps in that understanding. To develop effective prompts, we followed an iterative prompt engineering process with human validation at each step, focusing on relevance, specificity, and alignment with the curriculum’s context and goals.

In most instances, the GPT model generated accurate and reliable summaries. However, it struggled with identifying common misconceptions or gaps, occasionally producing hallucinated content. For example, in a simpler version of the prompt (zero-shot), GPT confused the curriculum’s focus on rainfall and runoff with the broader water cycle, incorrectly suggesting that teachers should emphasize the water cycle, which was not part of the curriculum.

To address these issues, we refined the prompt to include additional context, such as the agent’s role, the curriculum’s goals, relevant NGSS crosscutting concepts, and specific assessment questions using guidance from literature~\cite{white2023promptpatterncatalogenhance}. For questions targeting scientific knowledge (e.g., definitions of rainfall, runoff, and absorption), we included correct example responses as few-shot instances to help guide the model. The improved prompt was then tested using real classroom data from a previous implementation of the same curriculum. Overall, this human-in-the-loop design ensured that the GenAI component would serve as a useful instructional aid rather than a black-box solution~\cite{cohn2024chain}.

\section{Method}

\subsection{Participants}
Two 6th-grade teachers (1 male, 1 female) from an urban public school with over 34 years of STEM teaching experience participated in the study. Both teachers had previously implemented the Earth science curriculum in their classrooms and were actively involved in the co-design process during professional development sessions, making them well-suited for providing informed feedback. They provided informed consent to participate in the study. Teacher 1 implemented the Earth science curriculum across 3 class sections, while Teacher 2 taught one section. All study procedures and teacher interviews were approved by the Vanderbilt University IRB Committee.

\subsection{Data sources and Analysis}

We conducted semi-structured interviews with the two participating teachers three times during the implementation to evaluate the potential of the LearnLens dashboard. During each session, a researcher guided the teachers through the dashboard using a structured walkthrough; the teachers did not interact with the dashboard independently.

The interviews focused on two key areas: (1) teachers’ current data practices, including the types of reports or dashboards they typically use, and (2) feature-specific feedback on the LearnLens dashboard, including sample responses, word clouds, bar graphs, and AI-generated summaries and insights. Teachers were encouraged to share suggestions for improvement and reflect on the potential instructional use of each feature.

All sessions were audio- and screen-recorded. This paper analyzes the interview when the teachers were introduced to the dashboard for the first time. The 1-hour-long audio recording was transcribed and analyzed thematically, using a deductive coding scheme aligned with the interview structure\cite{deterding2021flexible}, sample codes include sense-making of presented analysis, clarifying analysis methods, connection with classroom experience, perception of features, and instructional modification. The major themes that emerged are discussed in the next section.
\section{Results and Findings from Teacher Interviews}

\textbf{Current Data Practices:} The teachers reported using a range of tools in their classroom instruction. Teacher 1 relied on more traditional methods, such as paper-based exit tickets, whereas Teacher 2 incorporated several digital tools, including Google Forms, Schoology (a Learning Management System), 
and EdPuzzle. Despite the variety of tools, both teachers emphasized a standard limitation, i.e., these platforms do not automatically generate summaries or insights. For instance, Teacher 2 said, \textit{``None of those things do anything except just organize it in a place. They don't give you any feedback.''}.

\noindent
\textbf{Dashboard Features Evaluation:}
Teachers appreciated the variety of visualizations we provided on the dashboard, particularly the word clouds for questions related to the definitions of scientific phenomena. Teacher 1 remarked, \textit{``I love word clouds!''}. Additionally, they found the bar graphs and filtering feature helpful, as these tools allowed them to examine students' understanding across different class sections. Displaying actual student responses verbatim was valuable because it enabled them to see exactly what students had written. The AI-generated misconceptions were also recognized as a beneficial feature: 
\begin{quote}
    \textit{Teacher 1: (Reading from screen) One student described absorption as involving electrons, which may indicate confusion with chemical processes like reactions or bonding. They don't even know about reactions or bonding. \\
    Teacher 2: So that, this misconception piece is super helpful here. 
    }
\end{quote}
\noindent
\textbf{Connection with Classroom Instruction \& Instructional Modifications:} 
While reviewing the dashboard, teachers could relate trends in student responses to the classroom discussions and shared how they would adjust their future instruction. For instance, Teacher 1 saw the word cloud as an accessible visualization and proposed printing and displaying it in the classroom to prompt student reflection (\textit{Teacher 1: ``seriously, we need to take some of these, if we can print them out and put the question with them, and just put them in various places around the classroom.''})

The AI-generated insights encouraged teachers to focus more on reinforcing vocabulary, particularly with key terms like ``absorption" and ``runoff," which are central components of the science curriculum. 
As Teacher 1 noted while reading from the screen: ``\textit{There are instances of incorrect spelling or uses of terms such as `absorps' or `seps', which could lead to misconceptions and misunderstandings or difficulty with terminology. `Absorption' is a hard word for kids, and we probably need to put it up on our vocabulary.}''

\noindent
\textbf{Trust in AI-generated Insights:}
When reviewing the AI summaries with the teachers, the researchers asked them whether they trusted the AI-generated insights, noting that the LLMs can occasionally produce inaccurate or unsupported information. Teachers expressed confidence in AI when it summarized student responses or identified gaps based on those responses. However, they were skeptical of AI-generated suggestions that went beyond the data, such as recommendations for modifying lesson plans, which should draw from broader instructional knowledge. Thus, while the teachers trusted the descriptive capabilities of the AI, they were cautious about its prescriptive use:
\begin{quote}
    \textit{
    Teacher 2: Well, in the place, I would come up with the trust pieces like when it gave suggestions on how to help misconceptions. I would have to see what they would say there. [Researcher: Oh, okay!] ... it is telling us what they got from the data. But in that case, if you're asking for a prescription of what to do that's coming, that wouldn't be coming from this data, so that would be a different.\\
    Teacher 1: But all this is from the data, so...\\
    Teacher 2: So that's trustable for me. [Teacher 1: Right!] It's not the same as just getting stuff like you're asking it for a prescription of what to do when this and this.}
\end{quote}

    


\section{Discussion and Future Work}
Our findings demonstrate that GenAI-enhanced dashboards can help teachers make sense of students' open-ended responses in exploratory learning environments. Both teachers emphasized the need for such dashboards as the current instructional support tools lack summary or feedback insights. When shown the dashboards' functionalities, the teachers valued its modular design, visualizations (e.g., word cloud, bar chart), and especially the AI-generated insights that highlighted common gaps or misconceptions in students' understanding. These features are well-aligned with several design principles proposed by Poh et al. (2023) ~\cite{poh_design_2023} for developing effective teacher-facing dashboards, such as simpler design, glanceable summaries, modular design, supporting sense-making and using real student data for evaluations. 

Interviews, revealed that teachers could connect the analysis with their own instruction, using the insights to inform future instruction. They found visualizations, like word clouds, accessible for the students and decided to display them in the classroom. While teachers trusted summaries grounded in student data, they were cautious about prescriptive suggestions extending beyond the data, such as instructional recommendations. The findings resonate with the topology outlined by Campos et al. ~\cite{campos_making_2021}. In the \textit{Emotional} dimension, teachers' response was \textit{Satisfactory} for conventional and \textit{Positive} for the AI-driven insights. In the \textit{Analytical} dimension, teachers could \textit{Recall} events from classes and \textit{Attribute} them to patterns in the data. In the \textit{Intentional} dimension, we could observe \textit{Action Intention} and \textit{Planning} of instructional adaptations. These results align with previous work, which shows that interactive technologies can support instructional adaptation by surfacing patterns in student understanding \cite{moed-abu_raya_teachers_2024,zhai_design_2024}.

The study reported in this paper is limited to a single interview with two teachers. Future work will incorporate additional data sources and longitudinal analysis to gain a deeper understanding of the sustained impact of AI-powered feedback on instruction. While there are growing efforts to integrate AI tools into educational practice \cite{price2025generative}, fewer studies evaluate the design with teacher feedback. This paper contributes to addressing this gap, providing evidence of teachers' positive experiences with AI-powered dashboards that offer timely, data-driven feedback. Additionally, beyond supporting instructional adaptation, teacher-facing dashboards, like LearnLens, may also encourage the adoption of open-ended curricula in the first place. By offering visibility into student thinking, such tools can make these pedagogical practices more approachable, especially for teachers who are less confident with them. In this way, LearnLens can function not just as a support tool during enactment but also as a catalyst for broader curricular change. Overall, this work highlights the value of simple yet timely insights that can effectively support teachers in addressing emerging classroom needs.


\bibliographystyle{splncs04}
\bibliography{sample-base}

\end{document}